\documentclass[a4paper]{article}

\newlength{\cw}
\def\nad#1{\mbox{\smash{\oalign{$#1$\crcr\hidewidth$\mathchar"017E$
\hidewidth}}}}
\def\arraystretch{1.2}

\begin{document}

 \bibliographystyle{unsrt}
 \title{Quantization of the tachyonic field and the preferred frame}
 \author{Jakub Rembieli\'nski\thanks{This work is supported under the
{\L}\'od\'z University grant no.\ 457.}\\
Katedra Fizyki Teoretycznej, Uniwersytet {\L}\'odzki\\
ul.~Pomorska 149/153, 90--236 {\L}\'od\'z, Poland\/%
 \thanks{{\it E-mail address\/}: jaremb@mvii.uni.lodz.pl}}
\date{}
 \maketitle

 \begin{abstract}
 A consistent quantization scheme for imaginary-mass field is proposed.
It is related to an appropriate choice of the synchronization procedure
(definition of time), which guarantee an absolute causality in agreement
with Lorentz covariance. In that formulation a possible existence of
field excitations (tachyons) distinguish an inertial frame (tachyon
privileged frame of reference) {\em via\/}  spontaneous breaking of the
so called synchronization group. In this scheme relativity principle is
broken but Lorentz symmetry is exactly preserved in agreement with local
properties of the observed world.
 \end{abstract}

 \section{Introduction}
 Some attention in the literature over last decades, related to the
question  of existence of faster-than-light particles, has been lacking
in view of the  apparent conflict with the causality principle.
Irrespective of an attempt  to reconcile the notion of superluminal
objects with causality on the  classical and/or semiclassical level
\cite{Rec1,OR,Rec2}, it is commonly  believed that there is no
respectable tachyonic quantum field theory at  present \cite{KK2}.

 However, in the last time we observe a return of interest in tachyons.
This  is related to some recent experimental data \cite{PDG,Ass,GR}
indicating that the square of the electron and muon neutrino masses seem
to be negative. Therefore the hypothesis that the neutrinos might
possible be a fermionic tachyons is now under the consideration
\cite{CHK,GMMR,Kos,CKPG,CK,Rem:neu}.

 On the other hand the admittance of space-like four-momentum eigenstates
can possibly extend quantum field theory by the weakness
of the spectral condition. Furthermore non-localizablity of tachyonic
modes may moderate QFT divergences. It is also noticeable that a
tachyonic condensate is an immanent point of superstring models
\cite{KS,KP,CG,DS,Tse,MM}.

 This paper is motivated by the problems mentioned above. Here we
propose a consistent quantization of a scalar imaginary-mass field. Our
quantization scheme is related to a nonstandard procedure of
synchronization of clocks introduced in \cite{Tan,Cha1,Cha2,Rem1} and a
new form of realization of Lorentz symmetry proposed in \cite{Rem1}.
This procedure allows us to introduce the notion of a coordinate time
appropriate to the definition of the universal notion of causality in
agreement with Lorentz covariance. The main results can be summarized as
follows:
 \begin{itemize}
 \item The relativity principle and the Lorentz covariance are
formulated in the framework of a nonstandard synchronization scheme
(the Chang--Thangherlini (CT) scheme). The absolute causality holds
for all kinds of events (time-like, light-like, space-like).
 \item For bradyons and luxons our scheme is fully equivalent to the
standard formulation of special relativity.
  \item For tachyons it is possible to formulate covariantly proper
initial conditions and the time development.
 \item There exists a (covariant) lower bound of energy for tachyons;
in terms of the contravariant zero-component of the four-momentum this
lower bound is simply zero.
 \item The paradox of ``transcendental'' tachyons apparent in the
standard approach disappears.
 \item Tachyonic field can be consistently quantized using the CT
synchronization scheme.
 \item Tachyons distinguish a preferred frame {\em via\/} mechanism of
the spontaneous symmetry breaking \cite{Rem2,Rem:tac}; consequently the
relativity principle is broken, but the Lorentz covariance (and
symmetry) is preserved. The preferred frame can be identified with the
cosmic background radiation frame.
 \end{itemize}

 The logical sequence of this paper is the following:
 \begin{enumerate}
 \item By means of  some freedom in the special relativity, related to
the fact that only round-trip light velocity is measurable and must be
the constant $c$, we derive the most general realization of the Lorentz
group in the bundle of inertial frames.
 \item We select two distinguished synchronization conventions: the
Einstein--Poincar\'e one (standard choice) and the Chang--Tangherlini
one (non-standard choice).
 \item We show that both synchronizations are completely equivalent if
we consider subluminal and light signals, however only the second one
(CT) is in agreement with faster than light propagation. Thus the CT
synchronization can be used to a consistent description of tachyons.
 \item We formulate a consistent free field theory for scalar tachyonic
field in CT synchronization scheme.
 \end{enumerate}

 In the forthcoming article we give classification of unitary
Poincar\'e orbits in CT-synchronization as well as quantum field
theoretical description of fermionic tachyons with helicities
$\pm\frac{1}{2}$ (see also \cite{Rem:neu}).

\section{Preliminaries}\label{pre}
 As is well known, in the standard framework of the special relativity,
space-like geodesics do not have their physical counterparts. This is an
immediate consequence of the assumed causality principle which admits
time-like and light-like trajectories only.

 In the papers by Terletsky \cite{Ter}, Tanaka \cite{Tnk},  Sudarshan
{\em et al.\/} \cite{BDS}, Recami {\em et al.\/}  \cite{Rec1,OR,Rec2}
and Feinberg \cite{Fei} the causality problem has been  reexamined and a
physical interpretation of space-like trajectories was  introduced.
However, every proposed solution raised new unanswered questions  of the
physical or mathematical nature \cite{T&M}. The difficulties are
specially frustrating on the quantum level \cite{KK,KK2,Nak}. It is
rather  evident that a consistent description of tachyons lies in a
proper extension  of the causality principle. Notice that interpretation
of the space-like world  lines as physically admissible tachyonic
trajectories favour the  constant-time initial hyperplanes. This follows
from the fact that only  such surfaces intersect each world line with
locally nonvanishing slope once  and only once. Unfortunately, the
instant-time hyperplane is {\em not a  Lorentz-covariant notion\/} in
the standard formalism, which is just the source of many troubles with
causality.

 The first step toward a solution of this problem can be found in the
papers by Chang \cite{Cha1,Cha2,Cha3}, who introduced four-dimensional
version of the Tangherlini transformations \cite{Tan}, termed the
Generalized Galilean Transformations (GGT). In \cite{Rem1} it was shown
that GGT, extended to form a group, are hidden (nonlinear) form of the
Lorentz group transformations with $SO(3)$ as a stability subgroup.
Moreover, a difference with the standard formalism lies in a nonstandard
choice of the synchronization procedure. As a consequence a
constant-time hyperplane is a covariant notion. In the following we will
call this procedure of synchronization the {\em Chang--Tangherlini
synchronization scheme}.

 It is important to stress the following two well known facts: (a)~the
definition of a coordinate time depends on the synchronization scheme
\cite{Rei,TV,Var}, (b)~synchronization scheme is a convention, because
no experimental procedure exists which makes it possible to determine
the one-way velocity of light without use of superluminal signals
\cite{Jam}. Notice that a choice of a synchronization scheme, different
that the standard one, {\em does not affect seriously the assumptions of
special relativity but evidently it can change the causality notion},
depending on the definition of the coordinate time.

 As it is well known, intrasystemic synchronization of clocks in their
``setting'' (zero) requires a definitional or conventional
stipulation---for discussion see Jammer \cite{Jam}, Sj\"odin
\cite{Sjo} (see also \cite{MS}). Really, to determine one-way light speed
it is necessary to use synchronized clocks (at rest) in their
``setting'' (zero)\footnote{Evidently, without knowledge of the one-way
light speed, it is possible to synchronize clocks in their rate only
\cite{And}.}. On the other hand to synchronize clocks we should know the
one-way light  velocity. Thus we have a logical loophole. In other words
no experimental procedure exists (if we exclude superluminal signals)
which makes possible to determine unambiguously and without any
convention the one-way velocity of light (for analysis of some
experiments see Will \cite{Wil}). Consequently, an {\em operational
meaning has the average value of the light velocity around closed paths
only}. This statement is known as the conventionality thesis \cite{Jam}.
Following Reichenbach \cite{Rei}, two clocks $\sf A$ and $\sf B$
stationary in the points $A$ and $B$ of an inertial frame are defined as
being synchronous with help of light signals if
$t_B=t_A+\varepsilon_{AB}(t'_A-t_A)$. Here $t_A$ is the emission time of
light signal at point $A$ as measured by clock $\sf A$, $t_B$ is the
reception-reflection time at point $B$ as measured by clock $\sf B$ and
$t'_A$ is the reception time of this light signal at point $A$ as
measured by clock $\sf A$. The so called synchronization coefficient
$\varepsilon_{AB}$ is an arbitrary number from the open interval
$(0,1)$. In principle it can vary from point to point. The only
conditions for $\varepsilon_{AB}$ follow from the requirements of
symmetry and transitivity of the synchronization relation. Note that
$\varepsilon_{AB}=1-\varepsilon_{BA}$. The one-way velocities of light
from $A$ to $B$ ($c_{AB}$) and from $B$ to $A$ ($c_{BA}$) are given by
\[ c_{AB}=\frac{c}{2\varepsilon_{AB}},\quad
 c_{BA}=\frac{c}{2\varepsilon_{BA}}. \]
 Here $c$ is the round-trip average value of the light velocity. In
standard  synchronization $\varepsilon_{AB}=\frac{1}{2}$ and
consequently $c=c_{AB}$  for each pair $A$, $B$.

 The conventionality thesis states that from the operational point of
view the  choice of a fixed set of the coefficients $\varepsilon$ is a
convention.  However, the explicit form of the Lorentz transformations
will be  $\varepsilon$-dependent in general. The question arises: Are
equivalent  notions of causality connected with different
synchronization schemes?  As we shall see throughout this work the
answer is {\em negative\/} if we  admit tachyonic world lines. In other
words, the causality requirement,  logically independent of the
requirement of the Lorentz covariance, can  contradict the
conventionality thesis and consequently it can prefer a  definite
synchronization scheme, namely CT scheme if an absolute causality is
assumed.

\section{The Chang--Tangherlini synchronization}\label{CTsyn}
 As was mentioned in Section \ref{pre}, in the paper by Tangherlini
\cite{Tan} a family of inertial frames in $1+1$ dimensional space of
events was introduced with the help of transformations which connect the
time coordinates by a simple (velocity dependent) rescaling. This
construction was generalized to the $1+3$ dimensions by Chang
\cite{Cha1,Cha2}. As was shown in the paper \cite{Rem1}, the
Chang--Tangherlini inertial frames can be related by a group of
transformations isomorphic to the orthochronous Lorentz group. Moreover,
the coordinate transformations should be supplemented by transformations
of a vector-parameter interpreted as the velocity of a privileged frame.
It was also shown that the above family of frames is equivalent to the
Einstein--Lorentz one; (in a contrast to the interpretation in
\cite{Cha1,Cha2}). A difference lies in another synchronization
procedure for clocks \cite{Rem1}. In the Appendix we derive  realization
of the Lorentz group given in \cite{Rem1} in a systematic way
\cite{Rem2}. An elegant discussion of particle mechanics in the CT
synchronization is given by Jaroszkiewicz \cite{Jar}.

 Let us start with a simple observation that the description of a family
of (relativistic) inertial frames in the Minkowski space-time is not so
natural. Instead, it seems that the geometrical notion of bundle of
frames is more natural. Base space is identified with the space of
velocities; each velocity marks out a coordinate frame. Indeed, from the
point of view of an observer (in a fixed inertial frame) all inertial
frames are labelled by their velocities with respect to him. Therefore,
in principle, to define the transformation rules between frames, we can
use, except of coordinates, also this vector-parameter, possibly related
to velocities of frames with respect to a distinguished observer.
Because we adopt Lorentz covariance, we can use a time-like
four-velocity $u_E$; subscript $E$ means Einstein--Poincar\'e
synchronization\footnote{In the papers by Chang \cite{Cha1,Cha2,Cha3} it
was used some kinematical objects with an unproper physical
interpretation \cite{SP,FN}. For this reason we should be precise in the
nomenclature related to different synchronizations.} (EP
synchronization) i.e.\ we adopt, at this moment, the  standard
transformation law for $u_E$
 \[ u_E'=\Lambda u_E \]
 where $\Lambda$ is an element of the Lorentz group $L$.

 Notice that a distinguishing of a preferred inertial frame is in full
agreement with {\em local\/} properties of the observed expanding world.
Indeed, we can fix a local frame in which the Universe appears spherically; it
can be done, in principle, by investigation of the isotropy of the Hubble
constant \cite{Wei}. It concides with the cosmic background radiation frame.
Thus it is natural to ask for a formalism incorporating locally Lorentz
symmetry and the existence of a preferred frame.

 Below we list our basic requirements:
 \begin{enumerate}
 \item Coordinate frames are related by a set of transformations
isomorphic to the Lorentz group ({\bf Lorentz covariance}).\label{i}
 \item The average value of the light speed over closed paths is
constant $(c)$ for all inertial observers ({\bf constancy of the
round-trip light velocity}).\label{ii}
 \item With respect to the rotations $x^0$ and $\vec{x}$ transform as
$SO(3)$ singlet and triplet respectively ({\bf isotropy}).\label{iii}
 \item Transformations are linear with respect to the coordinates ({\bf
affinity}).\label{iv}
 \item We admit an additional set of parameters $u_E$ ({\bf the base
space} for the bundle of inertial frames).\label{v}
 \end{enumerate}
 We see that assumptions labelled by \ref{i}--\ref{iv} are the standard
one, while \ref{v} is rather new one. In the following we consider also
two distinguished cases corresponding to the relativity principle and
absolute causality requirements respectively.

 \subsection{Lorentz group transformation rules in the standard and CT
synchronization}\label{Lor}
 According to our assumptions, transformations between two coordinate
frames $x^{\mu}$ and ${x'}^\mu$ have the following form
 \begin{equation}\label{1}
 x'(u_E')=D(\Lambda,u_E)x(u_E).
 \end{equation}
 Here $D(\Lambda,u_E)$ is a real (invertible) $4\times4$ matrix,
$\Lambda$ belongs to the Lorentz group and $u_{E}^{\mu}$ is assumed to
be a Lorentz four-vector, i.e.,
 \begin{equation}\label{2}
 u_E'=\Lambda u_E,\quad {u_E}^2=c^2>0.
 \end{equation}
 The physical meaning of $u_E^\mu$ will be explained later. It is easy
to verify that the transformations (\ref{1}--\ref{2}) constitute a
realization of the Lorentz group if the following composition law holds
 \begin{equation}\label{d2}
 D({\Lambda}_2,{\Lambda}_1u_E)D({\Lambda}_1,u_E)=
D({\Lambda}_2{\Lambda}_1,u_E).
 \end{equation}
 The explicit form of $D(\Lambda,u_{E})$ satisfying the assumptions
\ref{i}--\ref{v} is derived in the Appendix and it reads
 \begin{equation}\label{D}
 D(\Lambda,u_{E})=T(\Lambda u_{E})\Lambda T^{-1}(u_{E}),
 \end{equation}
 where
 \begin{equation}\label{13}
 T(u_E)=\left(\begin{array}{c|c}
 1&b(u^0_E)\vec u_E^{\rm T}\\[1ex]
 \hline
 0&I\end{array}\right).
 \end{equation}
 Here $b(u^{0}_{E})$ is a function of $u^{0}_{E}$; the superscript $^{\rm
T}$ denotes transposition. Thus the light velocity has the following
form (see eq.\ (\ref{A14}) in Appendix A)
 \begin{equation}\label{14}
 \vec c=c\vec n\left(1+b\vec u_E\vec n\right)^{-1},
 \end{equation}
so the Reichenbach coefficient reads
 \begin{equation}\label{15}
 \varepsilon(\vec n,\vec u_E)=\frac{1}{2}\left(1+b\vec u_E\vec n\right).
 \end{equation}

 It is evident that the function $b(u^0_E)$ distinguishes between
different synchronizations. Choosing $b(u^0_E)=0$ we obtain $\vec
c=c\vec n$, $\varepsilon=\frac{1}{2}$ and the standard transformation
rules for coordinates: $x'_{E}=\Lambda x_{E}$, where, as before the
subscript~$_{E}$ denotes EP-synchronization. On the other hand, if we
demand that the instant-time hyperplane $x^0={\rm constant}$ be an
invariant notion, i.e.\ that ${x'}^0={D(\Lambda, u_E)^0}_0 x^0$ so
${D(\Lambda, u_E)^0}_k=0$, then from eqs.\ (\ref{D}, \ref{13}) we have
 \begin{equation}\label{16}
 b(u^0_E)=-\frac{1}{u^0_E}.
 \end{equation}
 In the following we restrict ourselves to the above case defined by
eq.\ (\ref{16}). Notice that $\vec u_E/u^0_E$ can be expressed by a
three-velocity $\vec\sigma_E$
 \begin{equation}\label{17}
 \frac{\vec u_E}{u^0_E}=\frac{\vec\sigma_E}{c}
 \end{equation}
with $0\leq|\vec\sigma_E|<c$. Therefore
 \begin{equation}\label{18a}
 T(u_{E})=\left(\begin{array}{c|c}
 1&\displaystyle-\frac{\vec\sigma^{\rm T}_E}{c}\\[1ex]
 \hline
 0&I
 \end{array}\right).
 \end{equation}
 Thus we have determined by (\ref{2}, \ref{D}, \ref{18a}) the form of
the transformation law (\ref{1}) in this case. Now, according to our
interpretation of the freedom in the Lorentz group realization as the
synchronization convention freedom, there should exists a relationship
between $x^\mu$ coordinates and the Einstein--Poincar\'e coordinates
$x^{\mu}_{E}$. In fact, the matrix $T$ relates both synchronizations
{\em via\/} the formula
 \begin{equation}\label{D3}
 x=T(u_{E})x_{E}.
 \end{equation}
 Explicitly:
 \begin{equation}\label{D4}
 x^0=x^0_E-\frac{\vec\sigma_E}{c}\vec x_E,\qquad
 \vec x=\vec x_E.
 \end{equation}
 It is easy to check that $x_E$ transforms according to the standard law
i.e.\
 \begin{equation}\label{D5}
 x_E'=\Lambda x_E.
 \end{equation}
 Now, by means of eq.\ (\ref{D4}) we obtain analogous relations between
differentials
 \begin{equation}\label{D6}
 dx^0=dx^0_E-\frac{\vec\sigma_E}{c}\,d\vec x_E,\qquad
 d\vec x=d\vec x_E,
 \end{equation}
 and consequently interrelations between velocities in both
synchronizations; namely
 \begin{equation}\label{D7}
 \vec v=\frac{\vec v_E}{1-\displaystyle\frac{\vec
v_E\vec\sigma_E}{c^2}},
 \end{equation}
 \begin{equation}\label{D8}
 \vec v_E=\frac{\vec v}{1+\displaystyle\frac{\vec\sigma\vec v}{c^2}
\gamma^{-2}_0}.
 \end{equation}
 Here $\vec\sigma$ is the $\vec\sigma_E$ velocity in the CT
synchronization, i.e.,
 \begin{equation}\label{D9}
 \vec\sigma=\frac{\vec\sigma_E}{1-\left(\displaystyle\frac{\vec\sigma_E}{c}
\right)^2},
 \end{equation}
 while $\gamma_0$ is defined as
 \begin{equation}\label{D10}
 \gamma_0=\left[\frac{1}{2}\left(1+\sqrt{1+\left(\displaystyle\frac{2
\vec\sigma}{c}\right)^2}\right)\right]^{1/2}.
 \end{equation}
 In the following we use also the quantity $\gamma(\vec v)$ defined as
follows
 \begin{equation}\label{D11}
 \gamma(\vec v)=\left|\left(1+\frac{\vec\sigma\vec
v}{c^2}\gamma^{-2}_0\right)^2-\left(\frac{\vec
v}{c}\right)^{2}\right|^{1/2}.
 \end{equation}
 Now, taking into account eqs.\ (\ref{14}, \ref{16}, \ref{17}) we see
that the light velocity $\vec c$ in the direction of a unit vector $\vec
n$ reads
 \begin{equation}\label{D12}
 \vec c=\frac{c\vec n}{1-\displaystyle\frac{\vec
n\vec\sigma_E}{c}},
 \end{equation}
 i.e.\ in terms of $\vec\sigma$ (see eq.\ (\ref{D9}))
 \begin{equation}\label{D13}
 \vec c=\frac{c\vec n}{1-\displaystyle\frac{\vec n\vec\sigma}{c}
\gamma_0^{-2}},
 \end{equation}
 so
 \begin{equation}\label{D14}
 \varepsilon(\vec
n,\vec\sigma)=\frac{1}{2}\left(1-\frac{\vec n\vec\sigma}{c}
\gamma_0^{-2}\right).
 \end{equation}
 We call the synchronization scheme defined by the above choice of the
Reichenbach coefficients the {\em Chang--Tangherlini synchronization}.
In terms of $\vec\sigma$ or the four-velocity $u=T(u_{E})u_{E}$ (in the
CT synchronization) the matrix $T$ reads
 \begin{equation}\label{X}
 T(u)=\left(\begin{array}{c|c}
 1&-\displaystyle\frac{u^{0}\vec{u}^{\rm T}}{c^{2}}\\[1ex]
 \hline
 0&I
 \end{array}\right)=\left(\begin{array}{c|c}
 1&\displaystyle-\frac{\vec\sigma^{\rm T}}{c}\gamma_0^{-2}\\[1ex]
 \hline
 0&I
 \end{array}\right).
 \end{equation}

 Let us return to the transformation laws (\ref{1}) and (\ref{2}). By
means of the formulas (\ref{D}, \ref{D8}) and (\ref{X}) we can deduce
the explicit form of the Lorentz group transformations $D(\Lambda,u)$
expressed in terms of in the CT synchronization variables
\cite{Rem1,Rem2}. We give below the explicit form of the
transformation law
 \begin{equation}\label{...}
 x'=D(\Lambda,u)x,\qquad
 u'=D(\Lambda,u)u,
 \end{equation}
 where, for convenience, we use three-velocity
$\frac{\vec{\sigma}}{c}=\frac{\vec{u}}{u^{0}}$ instead of $u^{\mu}$.
 \begin{description}
 \item[Boosts] \hfil
 \begin{equation}\label{D15a}
 {x'}^0=\gamma x^0,
 \end{equation}
 \begin{equation}\label{D15b}
 {\vec x}'=\vec x+\frac{\vec V}{c}\left[\frac{\vec V\vec x}{c
\left(\gamma+\sqrt{\gamma^2+\left(\displaystyle\frac{\vec
V}{c}\right)^2} \right)}-\frac{\vec\sigma\vec
x}{c\gamma^2_0}-x^0\right]\gamma^{-1},
 \end{equation}
 \begin{equation}\label{D15c}
 {\vec\sigma}'=\vec\sigma\gamma^{-1}+\vec V\gamma^{-2} \left[\frac{\vec
V\vec\sigma}{c^2\left(\gamma+\sqrt{\gamma^2+
\left(\displaystyle\frac{\vec V}{c}\right)^2}\right)}
-\left(\frac{\vec\sigma}{c}\right)^2\gamma_0^{-2}-1\right].
 \end{equation}
 Here $\gamma=\gamma(\vec{V})$ has the form (\ref{D11}).
 \item[Rotations] \hfil
 \begin{equation}\label{D16a}
 {x'}^0=x^0,
 \end{equation}
 \begin{equation}\label{D16b}
 {\vec x}'=R\vec x,
 \end{equation}
 \begin{equation}\label{D16c}
 {\vec\sigma}'=R\vec\sigma.
 \end{equation}
 \end{description}
 It is easy to see that a vector $\vec{V}$ appearing in the
transformations rules (\ref{D15a}--\ref{D15c}) is the relative velocity
of the frame $x'$ with respect to $x$, measured in the CT
synchronization. Moreover, from (\ref{D15a}--\ref{D16c}) we can deduce
the meaning of the vector-parameter $\vec{\sigma}$; namely
$\vec{\sigma}$ is the velocity of a fixed (formally privileged) frame as
measured by an observer which uses the coordinates $x(\vec\sigma)$. The
four-velocity $u^{\mu}$ transforms like $x^{\mu}$, of course.

Notice that the matrix $D$ (eq.\ (\ref{1})) for Lorentz boosts reads
 \begin{equation}\label{D20}
 D(\vec{V},u)=\left(\begin{array}{c|c}
 \gamma & 0\\[1ex]
 \hline
 \displaystyle-\frac{\vec V}{c}\gamma^{-1} & I+ \displaystyle\frac{\vec
V\otimes\vec V^{\rm T}}
{c^2\gamma\left[\gamma+\sqrt{\gamma^2+\displaystyle\frac{\vec V}{c}^2}
\right]}-\frac{\vec V\otimes\vec\sigma^{\rm T}}{c^2\gamma\gamma^2_0}
 \end{array}\right).
 \end{equation}
 By means of (\ref{...}) it is easy to see that the bilinear form
$x^{\rm T}(u)g(u)x(u)$ with $g(u)=\bigl(T(u)\eta T^{\rm
T}(u)\bigr)^{-1}$, where $\eta$ is the Minkowski metric tensor is a
Lorentz group invariant. For completeness, we give also the explicit
form of the metric tensors $g_{\mu\nu}(u)$ and
$g^{\mu\nu}(u)$:
 \begin{equation}\label{D17}
 \left[g_{\mu\nu}(u)\right]=\left(\begin{array}{c|c}
 1 & \displaystyle\frac{u^{0}\vec{u}^{\rm T}}{c^{2}}\\[1ex]
 \hline
 \displaystyle\frac{u^{0}\vec{u}}{c^{2}} &
\displaystyle-I+\frac{\vec{u}\otimes\vec{u}^{\rm T}}{c^{4}}
(u^{0})^{2}
 \end{array}\right)=\left(\begin{array}{c|c}
 1 & \displaystyle\frac{\vec\sigma^{\rm T}}{c}\gamma_0^{-2}\\[1ex]
 \hline
 \displaystyle\frac{\vec\sigma}{c}\gamma_0^{-2}&
\displaystyle-I+\frac{\vec\sigma\otimes\vec\sigma^{\rm T}}{c^{2}}
\gamma_0^{-4}
 \end{array}\right),
 \end{equation}
 \begin{equation}\label{D18}
\left[g^{\mu\nu}(u)\right]=\left(\begin{array}{c|c}
 \left(\displaystyle\frac{u^{0}}{c}\right)^{2} &
\displaystyle\frac{u^{0}\vec{u}^{\rm T}}{c^{2}}\\[1ex]
 \hline
 \displaystyle\frac{u^{0}\vec{u}}{c^{2}} & -I
 \end{array}\right)=\left(\begin{array}{c|c}
 \gamma_0^{-2}&\displaystyle\frac{\vec\sigma^{\rm T}}{c}\gamma_0^{-2}\\[1ex]
 \hline
\displaystyle\frac{\vec\sigma}{c}\gamma_0^{-2}& -I
 \end{array}\right).
 \end{equation}
 From (\ref{D18}) it is evident that the configuration three-space is
the Euclidian one. Furthermore, the subset of transformations
(\ref{D15a}--\ref{D15c}) defined by the condition $\vec\sigma=0$ coincides
exactly
with the family of the Chang--Tangherlini inertial frames
\cite{Cha1,Cha2}.

\subsection{Causality and kinematics in the CT synchronization}\label{caus}
 In this subsection we discuss shortly differences and similarities of
kinematical descriptions in both CT and EP synchronizations. Recall that
in CT scheme {\em causality has an absolute meaning}. This follows from
the transformation law (\ref{D15a}) for the coordinate time:
$x^0$ is rescaled by a positive, velocity dependent, factor $\gamma$.
Thus this formalism extends the EP causality by allowing faster than
light propagation. It can be made transparent if we consider the
relation derived from eq.\ (\ref{D6})
 \begin{equation}\label{D19}
 \frac{dx^0}{dx^0_E}=1-\frac{\vec\sigma_E\vec v_E}{c^2}.
 \end{equation}
 For $|\vec v_E|\leq c$ we have $\frac{dx^0}{dx^0_E}>0$, whereas for
$|\vec v_E|>c$, $\frac{dx^0}{dx^0_E}$ can be evidently negative which
is a consequence of an inadequacy of the EP synchronization to
description of faster than light propagation. Notice that subluminal
(superluminal) signals in the EP synchronization remain subluminal
(superluminal) in the CT one too; indeed, as we see from eqs.\
(\ref{D7}, \ref{D8}, \ref{D12}, \ref{D13}) the rate of the corresponding
velocities in the same direction $\vec{n}$ reads
 \[ \frac{|\vec{v}|}{|\vec{c}|}=\left|\frac{\displaystyle\frac{v_E}{c}-
\frac{\vec{v}_{E}\vec{\sigma}_{E}}{c^2}}
{1-\displaystyle\frac{\vec{v}_{E}\vec{\sigma}_{E}}{c^2}}\right|<1\quad
\mbox{iff}\quad\left|\frac{v_E}{c}\right|<1. \]

 Let us consider in detail a space-like four-momentum $k^{\mu}$
transforming under eq.\ (\ref{...}). Now, our $k$ satisfy
 \begin{equation}\label{D22}
 k^{2}=g_{\mu\nu}(\vec{\sigma})k^{\mu}k^{\nu}=
 g^{\mu\nu}(\vec{\sigma})k_{\mu}k_{\nu}=-\kappa^2<0.
 \end{equation}
 Because velocity of a particle has direct physical meaning we solve the
tachyonic dispersion equation (\ref{D22}) by means of the evident
relations
 \begin{equation}
 k^{\mu}=\frac{\kappa}{c}w^{\mu}
 \end{equation}
 with $w^{2}=-c^{2}$, and
 \begin{equation}
 \frac{\vec{v}}{c}=\frac{d\vec{x}}{dx^0}=\frac{\vec{w}}{w^0}.
 \end{equation}
 Consequently the solution of eq.\ (\ref{D22}) reads
 \begin{equation}\label{D23a}
 k^{0}_{\pm}=\pm\kappa\gamma^{-1},
 \end{equation}
 \begin{equation}\label{D23b}
 \vec{k}_{\pm}=\pm\kappa\gamma^{-1}\frac{\vec{v}}{c},
 \end{equation}
 where $\gamma=\gamma(\vec{v})$ is given by eq.\ (\ref{D11}).

 Now, by means of (\ref{D17}) the covariant four-momentum $k_{\mu}$ has
the form
 \begin{equation}\label{D24a}
 k_{0\pm}=\pm\kappa\gamma^{-1}\left(1+\gamma^{-2}_{0}
\frac{\vec{\sigma}\vec{v}}{c^2}\right),
 \end{equation}
 \begin{equation}\label{D24b}
 \nad{k}_{\pm}=\pm \kappa\gamma^{-1}\left[-\frac{\vec{v}}{c}+
\frac{\vec{\sigma}}{c}\gamma_{0}^{-2}\left(
1+\gamma^{-2}_{0}\frac{\vec{\sigma}\vec{v}}{c^2}\right)\right].
 \end{equation}
 Recall that the generators of space-time translations are covariant, so
energy must be identified with $k_0$. To make a proper identification of
energy ($k_{0+}$ or $k_{0-}$), let us analyse the above formulas with
the help of convenient parameters $\xi$, $s$ and $\varepsilon$
 \begin{equation}\label{D25a}
 \xi=\frac{|\vec{v}|}{|\vec{c}|} \in(1,\infty),\quad\mbox{(for tachyons)},
 \end{equation}
 \begin{equation}\label{D25b}
 s=\frac{|\vec{c}|}{c}\in(\frac{1}{2},\infty),
 \end{equation}
 \begin{equation}\label{D25c}
 \varepsilon=\gamma_{0}^{-2}\frac{\sigma}{c}\in\langle0,1),
 \end{equation}
 where in the eqs.\ (\ref{D25a}--\ref{D25b}) $\vec{c}$ is assumed to
propagate in the direction of $\vec{v}$, i.e., $\vec{n}$ in eq.\
(\ref{D13}) is taken in the form $\vec{n}=\vec{v}/|\vec{v}|$. In terms
of $\xi$ and $s$
 \begin{equation}\label{D26}
 k_{0\pm}=\pm\kappa\frac{1+(s-1)\xi}{\sqrt{(\xi-1)[(2s-1)\xi+1]}}.
 \end{equation}
 We see that a proper choice for tachyon energy is $k_{0+}$; indeed
$k_{0+}$ has a lower bound. Moreover, this property is covariant because
$k^{0}_{+}$ is positive in that case and
$\varepsilon(k^{0}_{+})=1={\rm invariant}$, as follows from the
eq.~(\ref{D15a}). Notice also that the lowest, asymptotic value of
energy $k_{0+{\rm min}}=\kappa(s-1)/(2s-1)^{1/2}$, corresponding to the
lowest, asymptotic value $k^{0}_{+{\rm min}}=0$, depends only on the
light propagation characteristics in a given frame. Thus in the CT
synchronization, contrary to the EP one, tachyonic energy is bounded
from below. This fact is especially important because implies stability
on the quantum level. Furthermore, invariance of the sign of $k^0$
allows the covariant decomposition of the tachyon field on the creation
and annihilation part, so the Fock procedure can be applied.

 Finally, let us reexamine the problem of the so called
``transcendental'' tachyon. To do this, recall the transformation law
for velocities in the EP synchronization \cite{And}
 \begin{equation}\label{XX}
 {\vec v}'_E=\frac{\gamma_{E}\vec{v}_E+\vec{V}_E\left[\displaystyle
\frac{\vec{V}_E\vec{v}_E}{c^2}\left(\gamma_E+1\right)^{-1}-1\right]}
{\displaystyle 1-\frac{\vec{V}_E\vec{v}_E}{c^2}},
 \end{equation}
 where $\gamma_E=\sqrt{1-\left(\frac{\vec{V}_E}{c}\right)^2}$.

 We observe that the denominator of the above transformation rule can
vanish for $\left|\vec{v}_E\right|>c$; Thus a tachyon moving with
$c<\left|\vec{v}_E\right|<\infty$ can be converted by a finite Lorentz
map into a ``transcendental'' tachyon with
$\left|\vec{v}_E'\right|=\infty$. This discontinuity is an apparent
inconsistency of this transformation law; namely in the EP scheme {\em
tachyonic velocity space does not constitute a representation space for
the Lorentz group\/}! A technical point is that the space-like
four-velocity cannot be related to a three-velocity in this case.

 On the other hand, in the CT scheme, the corresponding transformation
rule for velocities follows directly from eqs.\ (\ref{D15a}--\ref{D15c})
and reads
 \begin{equation}\label{XXX}
 {\vec v}'=\gamma^{-1}\vec{v}+\gamma^{-2}\vec{V}\left[\displaystyle
\frac{\vec{V}\vec{v}}{c^2\left(\gamma+\sqrt{\gamma^2+\displaystyle
\frac{\vec{V}^2}{c^2}}
\right)}-\frac{\vec{\sigma}\vec{v}}{c^2}\gamma_{0}^{-2}-1\right],
 \end{equation}
 where $\gamma=\gamma(\vec{V})$. Contrary to eq.\ (\ref{XX}), the
transformation law (\ref{XXX}) is continuous, does not ``produce''
``transcendental'' tachyons and completed by rotations, forms (together
with the mapping $\vec{\sigma}\rightarrow{\vec\sigma}'$) a realization
of the Lorentz group and the relation of $\vec{v}$ to the four-velocity
is nonsingular.

 We end this section with a Table \ref{tI} summarizing our results.

 \begin{table}
 \renewcommand{\arraystretch}{1}
 \caption{Comparison of the descriptions of kinematics in the
Einstein--Poincar\'e and Chang--Tangherlini synchronization
schemes.\label{tI}}
 \begin{center}
 \scriptsize
 \begin{tabular}{clcc}\hline\hline
\multicolumn{2}{l}{Synchronization scheme $\longrightarrow$}&
\multicolumn{1}{c}{Einstein--Poincar\'e}&
\multicolumn{1}{c}{Chang--Tangherlini}\\
\multicolumn{2}{l}{Description of $\downarrow$}&&\\
 \hline\hline
\multicolumn{1}{c}{\parbox{.7\cw}{\centering BRADYONS \\ $k^2=\kappa^2$ \\
and LUXONS \\ $k^2=0$}}&&
\parbox{1.2\cw}{\raggedright Consistent causal kinematics, fully
equivalent to the CT description}&
\parbox{1.2\cw}{\raggedright Consistent causal kinematics, fully
equivalent to the EP description}\\
 \hline
&universal notion&no&yes\\ &of causality&&($\varepsilon(dx^0)={\rm inv}$)\\
 \cline{2-4}
&covariant initial&no&yes\\ &conditions&&($x^0={\rm const}$)\\
 \cline{2-4}
TACHYONS&invariant sign of $k^0$&no&yes\\
$k^2=-\kappa^2$&&&($\varepsilon(k^0)={\rm inv}$)\\
 \cline{2-4}
&covariant lower&no&yes\\
&bound of energy&($k_0\to-\infty$)&($k_{0{\rm min}}>-\infty$\\
&&&or $k^{0}_{{\rm min}}\geq0$)\\
 \cline{2-4}
&paradox of ``trans-&inconsistency&
consistent (continuous)\\
&cendental'' tachyons&(discontinuity)&picture\\
 \hline\hline
 \end{tabular}
 \end{center}
 \end{table}

\subsection{Synchronization group and the relativity principle}\label{syn}
 From the foregoing discussion we see that the CT synchronization
prefers a privileged frame corresponding to the value $\vec{\sigma}=0$
({\em relativistic ether\/} \cite{Rem2}). It is clear that if we forget
about tachyons such a preference is only formal; namely we can choose
each inertial frame as a preferred one.

 Let us consider two CT synchronization schemes, say $A$ and $B$, under
two different choices of privileged inertial frames, say $\Sigma_A$ and
$\Sigma_B$. Now, in each inertial frame $\Sigma$ two coordinate charts
$x_A$ and $x_B$ can be introduced, according to both schemes $A$ and $B$
respectively. The interrelation is given by the almost obvious relations
 \begin{equation}\label{X1a}
 x_B=T(u_{E}^{B})T^{-1}(u_{E}^{A})x_A,
 \end{equation}
 \begin{equation}\label{X1b}
 u_{E}^{B}=\Lambda_{BA}u^{A}_{E},
 \end{equation}
 where $u^{A}_{E}(u^{B}_{E})$ is the four-velocity of
$\Sigma_A(\Sigma_B)$ with respect to $\Sigma$ expressed in the EP
synchronization for convenience. $T(u_E)$ is given by the eq.\
(\ref{18a}). We observe that a set of all possible four-velocities $u_E$
must be related by Lorentz group transformations too, i.e.\
$\{\Lambda_{BA}\}=L_S$. Of course it does not coincide with our
intersystemic Lorentz group $L$. We call the group $L_S$ a
synchronization group \cite{Rem2,Rem:tac}.

 Now, if we compose the transformations (\ref{1}, \ref{2}) of $L$ and
(\ref{X1a}--\ref{X1b}) of $L_S$ we obtain
 \begin{equation}\label{X2}
 (\Lambda_S,\Lambda):\qquad
 \left.\begin{array}{ll}
 x'=T(\Lambda_S\Lambda u_E)\Lambda T^{-1}(u_E)x,\\
 u_{E}'=\Lambda_{S}\Lambda u_E
 \end{array}\right\}
 \end{equation}
with $\Lambda_S\in L_S, \Lambda\in L$.

 Thus the composition law for $(\Lambda_S,\Lambda)$ reads
 \begin{equation}\label{X3}
 (\Lambda_S',\Lambda')(\Lambda_S,\Lambda)=
(\Lambda_S'(\Lambda'\Lambda_S{\Lambda'}^{-1}),\Lambda'\Lambda).
 \end{equation}
 Therefore, in a natural way, we can select three subgroups:
 \[ L=\{(I,\Lambda)\},\quad L_S=\{(\Lambda_S,I)\},\quad
L_0=\{(\Lambda_{S},\Lambda^{-1}_{S})\}. \]
 By means of (\ref{X3}) it is easy to check that $L_0$ and $L_S$
commute. Therefore the set $\{(\Lambda_S,\Lambda)\}$ is simply the
direct product of two Lorentz groups $L_0\otimes L_S$. The intersystemic
Lorentz group $L$ is the diagonal subgroup in this direct product. From
the composition law (\ref{X3}) it follows that $L$ acts as an
authomorphism group of $L_S$.

 Now, the synchronization group $L_S$ realizes in fact the relativity
principle. In our language the relativity principle can be formulated as
follows: {\em Any inertial frame can be choosen as a preferred frame}.
What happens, however, when the tachyons do exist? In that case the
relativity principle is obviously broken: {\em If tachyons exist then
one and only one inertial frame \underline{must be} a preferred frame\/}
to preserve an absolute causality. Moreover, the one-way light velocity
becomes a real, measured physical quantity because conventionality
thesis breaks down. It means that the synchronization group $L_S$ is
broken to the $SO(3)_{u}$ subgroup (stability group of $u_E$); indeed,
transformations from the $L_S/SO(3)_{u}$ do not leave the causality
notion invariant. As we show later, on the quantum level we have to deal
with {\em spontaneous breaking\/} of $L_S$ to $SO(3)$.

 Notice, that in the real world a preferred inertial frames are
distinguished {\em locally\/} as the frames related to the cosmic
background radiation. Only in such frames the Hubble constant is
direction-independent.

\section{Quantization}\label{quant}
 As was mentioned in Section \ref{CTsyn}, the following two related
facts, true only in the CT synchronization, are extremaly important for
a proper quantization procedure; namely the invariance of the sign
$\varepsilon(k^0)$ of $k^0$ and the existence of a covariant lower
energy bound. They guarantee an invariant decomposition of the tachyonic
field into a creation and an annihilation parts and stability of the
quantum theory. Recall that the non-invariance of
$\varepsilon(k^{0}_{E})$ and the absence of a lower bound for $k_{0E}$
were the main reason why the construction of a quantum theory for
tachyons in the EP synchronization scheme was impossible
\cite{KK2,KK,Nak}. This section is related to the approach presented in
\cite{Rem2,Rem:tac}.

\subsection{Dispersion relation $k^2=-\kappa^2$}
 The dispersion relation $k^2=-\kappa^2$ has, in the CT synchronization, the
following form (see eq.\ (\ref{D18})):
 \[ \left(\frac{u^0}{c}\right)^2+2\frac{u^0}{c}k_0\frac{\vec u}{c}\nad{k}
-\nad{k}^2+\kappa^2=0 \]
 i.e.
 \[ \left(\frac{uk}{c}\right)^2-\left(
\sqrt{\left(\frac{\vec{u}\nad{k}}{c}\right)^2+\nad{k}^2-\kappa^2}\right)^2=0
 \]
 where $uk=u^\mu k_\mu={\rm inv}$. Therefore we have two solutions
 \begin{equation}\label{*}
 \frac{uk_\pm}{c}
=\pm\sqrt{\left(\frac{\vec u \nad k}{c}\right)^2+\nad{k}^2-\kappa^2}
\equiv\pm\frac{c}{u^0}\omega_k
 \end{equation}
 with $k_\pm=(k_{0\pm},\nad{k})$. thus
 \begin{equation}\label{**}
 k_{0\pm}=\frac{c}{u^0}\left(\frac{uk_\pm}{c}-\frac{\vec u\nad k}{c}\right)
=-\frac{\vec u}{u^0}\nad k\pm\left(\frac{c}{u^0}\right)^2\omega_k.
 \end{equation}
Notice that the contravariant $k^0=\frac{u^0}{c}\left(\frac{uk}{c}\right)$, so
evidently
\[ \varepsilon(k^0)=\varepsilon\left(\frac{uk}{c}\right)={\rm inv}; \]
moreover, by means of (\ref{*}), $k^0_\pm=\pm\omega_k$, so
$\varepsilon(k^0_\pm)=\pm1$.

\subsection{Local tachyonic field and its plane-wave
decomposition}\label{loc}
 Let us consider a hermitian, scalar field $\varphi(x,u)$
satysfying the corresponding Klein--Gordon equation with imaginary
``mass'' $i\kappa$, i.e.\
 \begin{equation}
\left(g^{\mu\nu}(u)\partial_{\mu}\partial_{\nu} -
\kappa^2\right) \varphi(x,u)=0. \label{Q1}
 \end{equation}
 Our field $\varphi$ is $u$-dependent because (\ref{Q1}) is assumed to
be valid for an observer in an inertial frame moving with respect to the
privileged frame with the velocity $\vec{\sigma}$. Now, as in the
standard case, let us consider the Lorentz-invariant measure
 \begin{equation}\label{Q2}
d \mu(k,u)= \theta(k^0) \delta(k^2+\kappa^2) d^4 k.
 \end{equation}
 Notice, that $d\mu$ does not have an analog in the EP synchronization
because of non-invariance of the sign of $k_{E}^{0}$ in the EP case.

The Heaviside step function $\theta(k^0)$ guarantees the positivity of
$k^0$ and the lower bound of energy $k_0$ while $\delta(k^2+\kappa^2)$
projects on the $\kappa^2$-eigenspace of the d'Alembertian
$g^{\mu\nu}\partial_{\mu}\partial_{\nu}$. For this reason we can expand
invariantly the field $\varphi$ into the positive and negative
frequencies with respect to $k^0$
 \begin{equation}
 \varphi(x,u)=\frac{1}{(2\pi)^{3/2}}\int
d\mu(k,u)\left(
e^{ikx}\,a^{\dagger}(k,u)+e^{-ikx}\,a(k,u)\right).
 \end{equation}
 Integrating with respect to $k_0$ we obtain
 \begin{equation}\label{23}
 \varphi(x,u)=\frac{1}{(2\pi)^{3/2}}
\int_{\Gamma}\frac{d^3\nad{k}}{2\omega_k}
\left(e^{ik_{+}x}\,a^{\dagger}(k_{+},u)+
e^{-ik_{+}x}\,a(k_{+},u)\right).
 \end{equation}
 Here $k_{+}x=k_{0+}x^0+\nad{k}\vec{x}$ is given in (\ref{**}).
 The integration range $\Gamma$ is
determined by the constraint $k^2=-\kappa^2$, namely
 \begin{equation}\label{Q3}
 |\nad{k}|\geq\kappa\left(1+\left(\left(\frac{c}{u^0}\right)^{2}-1\right)
\left(\frac{\vec{u}\nad{k}}{|\vec{u}||\nad{k}|}\right)^2\right)^{-1/2},
 \end{equation}
 i.e., values of $\nad{k}$ lie outside the oblate spheroid with half-axes
$\kappa$ and $\kappa\frac{u^0}{c}$. Note that $\Gamma$ is
invariant under the inversion $\nad{k}\rightarrow-\nad{k}$.

 For the operators $a$ and $a^{\dagger}$ we postulate the canonical
commutation rules
 \begin{equation}\label{25a}
 \left[a(k_{+},u),a(p_{+},u)\right]
=\left[a^{\dagger}(k_{+},u),a^{\dagger}(p_{+},u)\right]
= 0,
 \end{equation}
 \begin{equation}\label{25b}
 \left[a(k_{+},u),a^{\dagger}(p_{+},u)\right]
=2\omega_{k}\delta(\nad{k}-\nad{p}).
 \end{equation}
 The vacuum $\left | 0 \right >$ is assumed to satisfy the conditions
 \begin{equation}\label{26}
 \left<0|0\right>=1\quad \mbox{and}\quad a(k_{+},u) \left |0 \right >
= 0.
 \end{equation}
 By the standard procedure, using eq.\ (\ref{23}), we obtain the
commutation rule for $\varphi(x,u)$ valid for an arbitrary
separation
 \begin{equation}\label{27}
 \left [ \varphi(x,u), \varphi(y,u) \right ] = -
i \Delta(x - y,u),
 \end{equation}
 where the analogon of the Schwinger function reads
 \begin{equation}
 \Delta(x,u) = \frac{- i}{(2\pi)^{3}} \int d^{4}k\,
\delta(k^{2} + \kappa^{2})\, \varepsilon(k^{0})\, e^{ikx}. \label{28}
 \end{equation}
 It is remarkable that $\Delta$ does not vanish for a space-like
separation which is a direct consequence of the faster-than-light
propagation of the tachyonic quanta. Moreover $\Delta(x,u)|_{x^{0}=0}=0$
and therefore no interference occurs between two measurements of
$\varphi$ at an instant time. This property is consistent with our
interpretation of instant-time hyperplanes as the initial ones.

 Now, because of the absolute meaning of the arrow of time in the CT
synchronization we can introduce an invariant notion of the time-ordered
product of field operators. In particular the tachyonic propagator
 \[ {\Delta}_{F}(x-y,u)=-i\left<0\right|
T(\varphi(x,u)\,\varphi(y,u))\left|0\right> \]
 is given by
 \begin{equation}\label{29}
 {\Delta}_{F}(x,u)=-\theta(x^0)\,
{\Delta}^{-}(x,u)+\theta(-x^0)\,{\Delta}^{+}(x,u)
 \end{equation}
 with
 \begin{equation}\label{Q4}
{\Delta}^{\pm}(x,u)=\frac{\mp i}{(2\pi)^3} \int
d^4k\,\theta(\pm k^0)\delta(k^2+\kappa^2)e^{ikx}.
 \end{equation}
 The above singular functions are well defined as distributions on the
space of ``well behaved'' solutions of the Klein--Gordon equation
(\ref{Q1}).  The role of the Dirac delta plays the generalized function
 \begin{equation}\label{30}
{\delta}^{4}_{\Gamma}(x-y)=\frac{1}{(2\pi)^3}\delta(x^0-y^0)
\int_{\Gamma}d^3\nad{k}\,e^{i\nad{k}(\vec{x}-\vec{y})}.
 \end{equation}
 The above form of ${\delta}^{4}_{\Gamma}(x)$ express impossibility of
the localization of tachyonic quanta. In fact, the tachyonic field does
not contain modes with momentum $\nad{k}$ inside the spheroid defined in
eq.\ (\ref{Q3}). Consequently, by the Heisenberg uncertainty relation,
an exact localization of tachyons is impossible.

 Note also that
 \[
\partial^0\Delta(x-y,u)\delta(x^0-y^0)={\delta}^{4}_{\Gamma}(x-y)
 \]
 so the equal-time canonical commutation relations for $\varphi(x,u)$
and its conjugate momentum $\pi(x,u)= \partial^0\varphi(x,u)$ have the
correct form
 \begin{equation}\label{q25a}
 \delta(x^0-y^0)\left[\varphi(x,u),\varphi(y,u)\right]=
\delta(x^0-y^0)\left[\pi(x,u),\pi(y,u)\right]=0,
 \end{equation}
 \begin{equation}\label{q25b}
 \left[\varphi(x,u),\pi(y,u)\right]\delta(x^0-y^0)=
i{\delta}^{4}_{\Gamma}(x-y)
 \end{equation}
 as the operator equations in the space of states.

 To do the above quantization procedure mathematically more  precise, we
can use wave packets rather than the plane waves. Indeed, with a help of
the measure (\ref{Q2}) we can define the Hilbert space
$H^{+}_{u}$ of one particle states with the scalar product
 \begin{equation}\label{C2}
 (f, g)_{u} = \int d\mu(k,u)\,
f^{\ast}(k,u)\, g(k,u)<\infty.
 \end{equation}
 Now, using standard properties of the Dirac delta we deduce
 \begin{equation}\label{XX3}
 (f,g)_{u}=\int_{\Gamma}\frac{d^3\nad{k}}{2\omega_k}
f^{*}(k_+,u) g(k_+,u).
 \end{equation}
 It is remarkable that for $\xi\rightarrow\infty$ (see eqs.\
(\ref{D25a}--\ref{D26})), $\omega_k\rightarrow 0$, so to preserve
inequality $\|f\|^{2}_{u}<\infty$, the wave packets $f(k_+,u)$ rapidly
decrease to zero with $\xi\rightarrow\infty$. This means physically that
probability of ``momentum localization'' of a tachyon in the infinite
velocity limit is going to zero in agreement with our intuition. As
usually we introduce the smeared operators
 \begin{equation}\label{C3}
 a(f,u) = (2\pi)^{-3/2}\int d\mu(k,u)\,
a(k,u)f^{\ast}(k,u)
 \end{equation}
 and the conjugate ones. The canonical commutation rules
(\ref{25a}--\ref{25b}) take the form
 \begin{equation}\label{C4a}
 \left[a(f,u), a(g,u)\right] =
\left[a^{\dagger}(f,u), a^{\dagger}(g,u)\right] =
0,
 \end{equation}
 \begin{equation}\label{C4b}
 \left[a(f,u), a^{\dagger}(g,u)\right] = (f,g)_{u}.
 \end{equation}
 We have also $a(f,u)\left|0\right> = 0$ and
$\left<f,u|g,u\right>=\left(f,g\right)_{u}$, where
$\left|f,u\right>=a^{\dagger}(f,u)\left|0\right>$. Let us discuss the
implementation of the intersystemic Lorentz group $L$ on the quantum
level. According to our assumption of scalarity of
$\varphi(x,u)$
 \begin{equation}\label{A}
 L\ni\Lambda:\;
\varphi'(x',u')=\varphi(x,u).
 \end{equation}
 where $x'$ and $u'$ are given by (\ref{...}). The
transformation law should be realized by a representation $U(L)$ as
follows
 \begin{equation}
 U(\Lambda)\varphi(x,u)U^{-1}(\Lambda)=\varphi(x',u'),
 \end{equation}
 i.e.,
 \begin{equation}
 U(\Lambda)a(k,u)U^{-1}(\Lambda)=a(k',u')
 \end{equation}
 and
 \begin{equation}\label{E}
 U(\Lambda)\left|0\right>=\left|0\right>.
 \end{equation}
 Therefore the wave packets must satisfy the scalarity condition
(\ref{A}) i.e.,
 \begin{equation}
 f'(k',u')=f(k,u).
 \end{equation}
 It follows that the family $\{U(\Lambda)\}$ forms an unitary orbit of
$L$ in the bundle of the Hilbert spaces $H^{+}_{u}$; indeed we see that
 \begin{equation}
 \left(f',g'\right)_{u'}=\left(f,g\right)_{u}.
 \end{equation}
 Summarizing, the Lorentz group $L$ is realized by a family of unitary
mappings in the following bundle of Hilbert spaces
 \begin{itemize}
 \item[] $H_0$ (vacuum);
 \item[] $\bigcup_{u}H^{+}_{u}$ (bundle of one-particle spaces of
states);
 \item[] $\bigcup_{u}H^{+}_{u}\otimes H^{+}_{u}$ (bundle of
two-particle spaces of states);
 \item[] \dotfill\ etc.\ \hspace*{\fill}
 \end{itemize}
 i.e.\ $H^{+}=H_0\oplus\bigl(\bigcup_{u}H^{+}_{u}\bigr)
\oplus\biggl(\bigcup_{u}\bigl(H^{+}_{u}\otimes H^{+}_{u}\bigr)\biggr)
\oplus\dots$ etc.\ with the base space as the velocity space
($u$-space). Now we introduce wave-packet solutions of the Klein--Gordon
equation {\em via\/} the Fourier transformation
 \begin{equation}
 {\cal F}(x,u) = (2
\pi)^{-3/2} \int d\mu(k,u)\, f(k,u)e^{-ikx}.
 \end{equation}
 In terms of these solutions the scalar product (\ref{C2}) reads
 \begin{equation}\label{C6}
 ({\cal F},{\cal G})_{u} = - i \int d^{3}\vec{x}\, {\cal
F}^{\ast}(x,u)\stackrel{\leftrightarrow}{\partial}^0 {\cal
G}(x,u).
 \end{equation}
 It is easy to see that for an orthonormal basis
$\{\Phi_{\alpha}(x,u)\}$ in $H^{+}_{u}$ the completeness
relation holds
 \begin{equation}\label{C7}
 \sum_{\alpha} \Phi^{\ast}_{\alpha}(x,u) \Phi_{\alpha}(y,u) = i
\Delta^{+}_{T}(x - y,u),
 \end{equation}
 where $\Delta^{+}$ has the form (\ref{Q4}) and it is the reproducing
kernel in $H^{+}_{u}$ i.e.\
 \[ (i \Delta^{+}(x,u), \Phi)_{u} = \Phi(x,u). \]
 Finally, translational invariance implies the following, almost
standard, form of the four-momentum operator
 \begin{equation}\label{XXX1}
P_{\mu}=\int d\mu(k,u)\,k_{\mu} a^{\dagger}(k,u)
a(k,u).
 \end{equation}
 It is evident that the vacuum $\left| 0 \right>$ has zero four-momentum.
Furthermore, $P_{\mu}$ applied to one-particle state
$a^{\dagger}(k_+,u)\left|0\right>=\left|k_+,u\right>$ gives
 \begin{equation}\label{XXX4}
P_{\mu} \left|k_+,u\right>=k_{\mu+}\left|k_+,u\right>
 \end{equation}
 Notice that the vacuum $|0\rangle$ is stable, because of the covariant
spectral condition $k^{0}_{+}>0$. Thus we have constructed a consistent
quantum field theory for the hermitian, scalar tachyon field
$\varphi(x,u)$. We conclude, that a proper framework to do this is the
CT synchronization scheme.

\subsection{Spontaneous breaking of the synchronization group}\label{break}
 As we have seen in the foregoing section, the intersystemic Lorentz
group $L$ is realized unitarily on the quantum level. In this section we
will analyse the role of the synchronization group $L_S$ in our scheme.

 As was stressed in the Sec.\ \ref{syn}, if tachyons exist then one and
only one inertial frame is the preferred frame. In other words the
relativity principle is broken in this case: tachyons distinguish a
fixed synchronization scheme from the family of possible CT
synchronizations. Consequently, because all admissible synchronizations
are related by the group $L_S$, this group should be broken. To see this
let us consider transformations belonging to the subgroup $L_0$ (see
Sec.\ \ref{syn}). They are composed from the transformations of
intersystemic Lorentz group $L$ and the synchronization group $L_S$;
namely they have the following form (see eq.\ (\ref{X2}) and the
definition of $L_0$),
 \begin{equation}\label{ZZ}
 u'=u,\qquad x'=T(u)\Lambda^{-1}_{S}T^{-1}(u)x
\equiv\Lambda^{-1}_{S}(u)x.
 \end{equation}
 We search an operator $W(\Lambda)$ implementing (\ref{ZZ}) on the
quantum level; namely
 \begin{equation}\label{ZZZ}
 \varphi'(x,u)=W(\Lambda_{S})\varphi(x,u)W^{\dagger}
(\Lambda_S)=\varphi(x',u).
 \end{equation}
 This means that we should compare both sides of (\ref{ZZZ}) i.e.
 \[ \int d\mu(k,u)\left[ e^{ikx}a^{\prime\dagger}
(k,u)+e^{-ikx}a^{\prime}(k,u)\right] \]
 \begin{equation}\label{84}
 =\int d\mu(p,u)\left[ e^{ipx'}a^{\dagger}
(p,u)+e^{-ipx'}a(p,u)\right],
 \end{equation}
 where $x'$ is given by eq.\ (\ref{ZZ}), while, formally
 \begin{equation}\label{85}
 a'=WaW^{\dagger},\qquad a^{\prime\dagger}=Wa^{\dagger}W^{\dagger}.
 \end{equation}
 Taking into account the form of the measure $d\mu$ (eq.\ (\ref{Q2}))
and the fact that $\Lambda_{S}(u)$ does not leave invariant
the sign of $k^0$, after some calculations, we deduce the following
form of $W$:
 \begin{equation}\label{86a}
 a'(k,u)=\theta(k^{\prime 0})a(k',u)+
\theta(-k^{\prime 0})a^{\dagger}(-k',u),
 \end{equation}
 \begin{equation}\label{86b}
a^{\prime\dagger}(k,u)=\theta(k^{\prime 0})a^{\dagger}(k',u)+
\theta(-k^{\prime 0})a(-k',u),
 \end{equation}
 where $k'=\Lambda_S(u)k$.

 We see that formally unitary operator $W(\Lambda_S)$ is realized by the
Bogolubov-like transformations; the Heaviside $\theta$-step functions
are the Bogolubov coefficients. The form (\ref{86a}--\ref{86b}) of the
transformations of the group $L_0$ reflects the fact, that a possible
change of the sign of $k^0$ causes a different decomposition of the
field $\phi$ on the positive and negative frequencies. Furthermore it is
easy to check that the transformation (\ref{86a}--\ref{86b}) preserves
the canonical commutation relations (\ref{25a}--\ref{25b}).

 However, the formal operator $W(\Lambda_{S})$ realized in the ring of the
field operators, cannot be unitarily implemented in the space of states
in general; only if $\Lambda=\Lambda_{u}$ is an element of the stability
group $SO(3)_u$ of $u$ in $L_S$, it can be realized unitarily. This is
related to the fact that $\Lambda_{u}$ does not change the
sign of $k^0$ for any $k$. Indeed, notice firstly that for $\Lambda_{S}\in
L_{S}/SO(3)$, $W(\Lambda_{S})$ does not anihilate the vacuum
$\left|0\right>$. Moreover, the particle number operator
 \begin{equation}\label{87}
 N=\int d\mu(k,u) a^{\dagger}(k,u) a(k,u)
 \end{equation}
 applied to the ``new'' vacuum
 \begin{equation} \label{88}
 \left|0\right>'=W^{-1}\left|0\right>
 \end{equation}
 gives
 \begin{equation}\label{89}
 N\left|0\right>'=\delta^3(0)\int_{\Gamma}
d^3\nad{k}\,\theta(-(\Lambda_S(u)k_+)^0)\left|0\right>'.
 \end{equation}
 The right side of the above expression diverges like $\delta^6(0)$ for
any $\Lambda_S(u)\in L_S/SO(3)_u$. Only for the stability subgroup
$SO(3)_u\subset L_S$ vacuum remains invariant. Thus, a ``new'' vacuum
$\left|0\right>'$, related to an essentially new synchronization,
contains an infinite number of ``old'' particles. As is well known, in
such a case, two Fock spaces $H$ and $H'$, generated by creation
operators from $\left|0\right>$ and $\left|0\right>'$ respectively,
cannot be related by an unitary transformation\footnote{We can treat
(\ref{86a}--\ref{86b}), in some sense, as a quantum version of the
familiar {\em reinterpretation principle\/} \cite{Fei}. We find that the
reinterpretation principle cannot be unitarily implemented.}
($W(\Lambda_S)$ in our case). Therefore, we have deal with the so called
spontaneous symmetry breaking of $L_S$ to the stability subgroup
$SO(3)$. This means that physically privileged is only one realization
of the canonical commutation relations (\ref{25a}--\ref{25b})
corresponding to a vacuum $\left|0\right>$ defined by eq.\ (\ref{26}).
Such a realization is related to a definite choice of the privileged
inertial frame and consequently to a definite CT synchronization scheme.
Thus we can conclude that, on the quantum level, {\em tachyons
distinguish a preferred frame via spontaneous breaking of the synchrony
group}.

 To complete discussion, let us apply the four-momentum operator
$P_{\mu}$ to the new vacuum $\left|0\right>'$. As the result we obtain
 \begin{equation}\label{90}
 P_{\mu}\left|0\right>'=-\delta^3(0){{\Lambda_S}_\mu}^{\nu}(u)
\int_{\Gamma}d^3\nad{k}\,\theta(-(\Lambda_S(u)k_+)^0)k_{\nu}
\left|0\right>'.
 \end{equation}
 This expression diverges again like $\delta^7(0)$ for $\Lambda_S\in
L_S/SO(3)_u$. Therefore a transition to a new vacuum ($\equiv$ change of
the privileged frame) demands an infinite momentum transfer, i.e.\ it is
physically inadmissible. This last phenomenon supports our claim that
existence of tachyons is associated with spontaneous breaking of the the
synchronization group.  On the other hand it can be simply shown
\cite{Rem2} that a free field theory for standard particles (bradyons or
luxons), formulated in CT synchronization, is unitarily equivalent to
the standard field theory in the EP synchronization.

\subsection{The stability of vacuum}\label{vacuum}
 One of the serious defects of the standard approach to the tachyon field
quantization is apparent instability of the vacuum. The reason is that
relativistic kinematics admits in this case many-particle states with
vanishing total four-momentum. It is related directly to the fact that for
each (space-like) four-momentum, say $k^\mu_E$, the four-momentum
$-k^\mu_E$ with the opposite sign is kinematically admissible, because
there is no spectral condition $k^0_E>0$ for space-like $k^\mu_E$.

Notwithstanding, such a situation does not take place in the presented scheme,
because space-like four-momentum $k$ satisfies the invariant spectral
condition, $k^0>0$ in each inertial frame\footnote{Recall that in the
asymptotics $k^0\to0$ the wave packets decrease to zero (see remark below the
eq.\ (\ref{XX3}).}. Thus the sum of $k$ and $k'$ satisfies the same spectral
condition. In brief, we have exactly the same situation as in the case of the
time-like (or light-like) four-momenta under the invariant spectral condition,
$k^0_{\rm E}>0$. This means that in our scheme multiparticle states with
vanishing total four-momentum do not appear, {\it ergo\/} vacuum
$\left|0\right>$ cannot decay. For example, for two particle state
$\left|q_+=k_++p_+\right>\equiv \left|k_+\right> \otimes \left|p_+\right>$,
where $k_+$ and $p_+$ satisfy spectral condition, i.e., $k^0_+>0$, $p^0_+>0$,
we have the inequality $q^0_+>0$ (i.e., $q_+\neq0$), so there is no
vacuum-like state with the four-momentum $q=0$. Concluding, this theory is
stable.

\section{Conclusions}\label{concl}
 We can conclude that, contrary to the current opinion, it is posible to
agree the Lorentz covariance and symmetry with universal causality and
existence of a preferred frame. Moreover, a consistent quantization of
the tachyonic field in this framework is possible and it is closely
connected with the choice of an appropriate synchronization scheme. From
this point of view the Einstein--Poincar\'e synchronization is useless
in the tachyonic case. On the other hand, in a description of bradyons
and luxons only, we are free in the choice of a synchronization procedure.
For this reason we can use in this case CT-synchronization as well as
the EP one.

 The CT-synchronization, a natural one for a description of tachyons,
favourizes a reference frame (privileged frame). This preference is only
formal if tachyons do not exist. However, if they exist, then an
inertial reference frame is really (physically) preferred, what in fact
holds in the real world. As a consequence, the one-way light velocity
can be measured in this case and, in general, it will be
direction-dependent for a moving observer. Light velocity is isotropic
only in the privileged frame.  On the other hand we have in the observed
world a serious candidate to such a frame; namely frame related to the
background radiation. Moreover, the standard cosmological model and
related models predict, except of locally distinguished preferred
inertial frame \cite{Wei}, also an absolute time (radius of the
universe), so the absolute causality too.

Of course, indirect arguments are not decisive ones for the existence of
tachyons. An experimental evidence can be a decisive argument only.  For this
reason it is very interesting that there are experiments \cite{PDG,Ass,GR}
suggesting that the square of masses of electron and muon neutrinos are
negative by a few standard deviations, so their tachyonic nature should be
seriously taken into account. In the forthcoming paper (see also
\cite{Rem:neu}) it is shown that it is possible to construct finite
component fermionic tachyon field theory, resembling in some aspects
Weyl's two component theory for neutrino.

\appendix
 \section{Derivation of the Lorentz group transformation rules}\label{der}
 Let us derive the form of transformations between two coordinate
frames $x^{\mu}$ and ${x'}^\mu$
 \begin{equation}\label{A1}
 x'(u_E')=D(\Lambda,u_E)x(u_E),
 \end{equation}
 where $D(\Lambda,u_E)$ is a real (invertible) $4\times4$ matrix,
$\Lambda$ belongs to the Lorentz group and $u_{E}^{\mu}$ is assumed to
be a Lorentz four-vector, i.e.,
 \begin{equation}\label{A2}
 u_E'=\Lambda u_E,\quad {u_E}^2=c^2>0.
 \end{equation}
 The transformations (\ref{A1}--\ref{A2}) constitute a realization of
the Lorentz group if the following composition law holds
 \begin{equation}\label{D2}
 D({\Lambda}_2,{\Lambda}_1u_E)D({\Lambda}_1,u_E)=
D({\Lambda}_2{\Lambda}_1,u_E).
 \end{equation}
 Now {\em we demand that\/} $(x^\mu)\equiv(x^0,\vec{x})$ {\em transform
under subgroup of rotations as singlet + triplet\/} ({\em isotropy
condition\/}), i.e.\ for $R\in SO(3)$
 \begin{equation}\label{A3}
 \Omega\equiv D(R,u_E)=\pmatrix{1&0\cr 0&R}.
 \end{equation}
 From eqs.\ (\ref{A1}--\ref{D2}) we see that the identity and the
inverse element have the form
 \begin{equation}\label{A4a}
 I=D(I,u_E),
 \end{equation}
 \begin{equation}\label{A4b}
 D^{-1}(\Lambda,u_E)=D(\Lambda^{-1},\Lambda u_E).
 \end{equation}
 Using the familiar Wigner's trick we obtain that
 \begin{equation}\label{A5}
 D(\Lambda,u_E)=T(\Lambda u_E)\Lambda T^{-1}(u_E),
 \end{equation}
 where the real matrix $T(u_E)$ is given by
 \begin{equation}\label{A6}
 T(u_E)=D(L_{u_E},\tilde u_E)L_{u_E}^{-1}.
 \end{equation}
 Here $\tilde u_E=(c,0,0,0)$ and $L_{u_E}$ is the boost matrix:
$u_E=L_{u_E}\tilde u_E$. We use the following parametrization of the
matrix $L_{u_E}$
 \[ L_{u_E}=\left(\begin{array}{c|c}
 \displaystyle\frac{u^0_E}{c}
 &\displaystyle\frac{\vec u_E^{\rm T}}{c}\\[1ex]
 \hline
 \displaystyle\frac{\vec u_E}{c}
 &\displaystyle I+\frac{\vec u_E\otimes\vec u_E^{\rm T}}{c^2\left(1+
\displaystyle\frac{u^0_E}{c}\right)}
 \end{array} \right). \]

 Note that the transformations (\ref{A1}--\ref{A2}) leave the bilinear
form  $x^{\rm T}(u_E)\* g(u_E)\* x(u_E)$, where the symmetric
tensor $g(u_E)$ reads
 \begin{equation}\label{A7}
 g(u_E)=(T(u_E)\eta T^{\rm T}(u_E))^{-1},
 \end{equation}
 invariant. Here $\eta$ is the Minkowski tensor and the superscript
$^{\rm T}$ means transposition.

 Now we determine the matrix $T(u_E)$. To do this we note that under
rotations
 \[ T(\Omega u_E)=\Omega T(u_E)\Omega^{-1}, \]
 so the most general form of $T(u_E)$ reads
 \begin{equation}\label{A8}
 T(u_E)=\left(\begin{array}{c|c}
 a(u^0_E)&b(u^0_E)\vec u_E^{\rm T}\\[1ex]
 \hline
 d(u^0_E)\vec{u_E}& e(u^0_E)I+(\vec u_E\otimes\vec u_E^{\rm T}) f(u^0_E)
 \end{array}\right),
 \end{equation}
 where $a$, $b$, $d$, $e$ and $f$ are some functions of $u^0_E$.
Inserting eq.\ (\ref{A8}) into eq.\ (\ref{A7}) we can express the metric
tensor $g(u_E)$ by $a$, $b$, $d$, $e$ and $f$. In a three dimensional
flat subspace we can use an orthogonal frame (i.e.\
$(g^{-1})^{ik}=-\delta^{ik};$ $i,k=1,2,3$), so we obtain
 \begin{equation}\label{A9}
 e(u^0_E)=1,\quad d^2=(2-f\vec u_E^2)f.
 \end{equation}
 Furthermore, from the equation of null geodesics, $dx^{\rm T}g\,dx=0$,
we deduce that the light velocity $\vec c$ in the direction $\vec n$
($\vec n^2=1$) is of the form
 \begin{equation}\label{A10}
 \vec c=c\vec n\left(\sqrt{\alpha+\beta^2\vec u_E^2}-\beta\vec u_E\vec
n\right)^{-1},
 \end{equation}
 where $\alpha=a^2-b^2\vec u_E^2$, $\beta=ad-b(1+f\vec u_E^2)$. From
eq.\ (\ref{A10}) we see that the synchronization convention depends on
the functions $\alpha$ and $\beta$ only. Now, because $a$, $b$ and $d$
can be expressed as functions of $\alpha$, $\beta$ and $f$ and we are
interested in {\em essentially different synchronizations\/} only, we can
choose
 \begin{equation}\label{A11a}
 f=0,
 \end{equation}
 so
 \begin{equation}\label{A11b}
 d=0,\quad\beta=-b,\quad\alpha=a^2-b^2\vec u_E^2.
 \end{equation}
 Finally, from (\ref{A10}--\ref{A11b}) the average value of $|\vec c|$
over a closed path is equal to
 \[ \langle|\vec c|\rangle_{\mbox{\scriptsize cl.\ path}}=\frac{c}{a}. \]
 Because {\em we demand that the round-trip light velocity\/}
($\langle|\vec c|\rangle_{\mbox{\scriptsize cl.\ path}}=c$) {\em be
constant}, we obtain
 \begin{equation}\label{A12}
 a=1.
 \end{equation}
 Summarizing, $T(u_E)$ has the form
 \begin{equation}\label{A13}
 T(u_E)=\left(\begin{array}{c|c}
 1&b(u^0_E)\vec u_E^{\rm T}\\[1ex]
 \hline
 0&I\end{array}\right),
 \end{equation}
while the light velocity
 \begin{equation}\label{A14}
 \vec c=c\vec n\left(1+b\vec u_E\vec n\right)^{-1},
 \end{equation}
so the Reichenbach coefficient reads
 \begin{equation}\label{A15}
 \varepsilon(\vec n,\vec u_E)=\frac{1}{2}\left(1+b\vec u_E\vec n\right).
 \end{equation}

 In the special relativity the function $b(u^0_E)$ distinguishes between
different synchronizations. Choosing $b(u^0_E)=0$ we obtain $\vec
c=c\vec n$, $\varepsilon=\frac{1}{2}$ and the standard transformation
rules for coordinates: $x'_{E}=\Lambda x_{E}$, where, as before the
subscript~$_{E}$ denotes EP-synchronization. On the other hand, if we
demand that the instant-time hyperplane $x^0={\rm constant}$ be an
invariant notion, i.e.\ that ${x'}^0={D(\Lambda, u_E)^0}_0 x^0$ so
${D(\Lambda, u_E)^0}_k=0$, then from eqs.\ (\ref{A5}, \ref{A13}) we have
 \begin{equation}\label{A16}
 b(u^0_E)=-\frac{1}{u^0_E}.
 \end{equation}
 Notice that $\vec u_E/u^0_E$ can be expressed by a three-velocity
$\vec\sigma_E$
 \begin{equation}\label{A17}
 \frac{\vec u_E}{u^0_E}=\frac{\vec\sigma_E}{c}
 \end{equation}
with $0\leq|\vec\sigma_E|<c$. Therefore
 \begin{equation}\label{A18a}
 T(u_E)=\left(\begin{array}{c|c}
 1&\displaystyle-\frac{\vec\sigma^{\rm T}_E}{c}\\[1ex]
 \hline
 0&I
 \end{array}\right).
 \end{equation}
 Thus we have determined the form of the transformation law (\ref{A1})
in this case in terms of the EP four-velocity $u_{E}$.


 \end{document}